\documentclass[twocolumn,aps,prl,amssymb,amsmath,superscriptaddress,showpacs,floatfix]{revtex4}
\usepackage{times}
\usepackage{graphicx}
\usepackage{epstopdf}
\usepackage{color}\usepackage[usenames,dvipsnames]{xcolor}
\usepackage{ulem} % for strikeout
\usepackage{xspace}

\def\AT{\textit{A}$_{\mathrm{T}}$}
\def\AB{\textit{A}$_{\mathrm{B}}$}
\def\BT{\textit{B}$_{\mathrm{T}}$}
\def\BB{\textit{B}$_{\mathrm{B}}$}
\def\IV{\textit{I}-\textit{V}}
\def\dIdV{d\textit{I}/d\textit{V}}
\def\R3byR3{$\sqrt{3}\times\sqrt{3}$}
\def\EF{\textit{E}$_{\mathrm{F}}$}
\def\ED{\textit{E}$_{\mathrm{D}}$}

\begin{document}
\title{Visualizing Atomic-Scale Negative Differential Resistance in Bilayer Graphene} 

\author{Keun Su Kim}\affiliation{Advanced Light Source, E. O. Lawrence Berkeley National Laboratory, Berkeley, CA 94720, USA}

\author{Tae-Hwan Kim}\affiliation{Center for Low Dimensional Electronic Symmetry and Department of Physics, Pohang University of Science and Technology, Pohang 790-784, Korea}

\author{Andrew L. Walter}\affiliation{Advanced Light Source, E. O. Lawrence Berkeley National Laboratory, Berkeley, CA 94720, USA}\affiliation{Department of Molecular Physics, Fritz-Haber-Institut der Max-Planck-Gesellschaft, Faradayweg 4-6, 14195 Berlin, Germany}

\author{Thomas Seyller}\affiliation{Lehrstuhl f\"{u}r Technische Physik, Universit\"{a}t Erlangen-N\"{u}rnberg, Erwin-Rommel-Strasse 1, 91058 Erlangen, Germany}

\author{Han Woong Yeom}\email[yeom@postech.ac.kr]{}\affiliation{Center for Low Dimensional Electronic Symmetry and Department of Physics, Pohang University of Science and Technology, Pohang 790-784, Korea}

\author{Eli Rotenberg}\affiliation{Advanced Light Source, E. O. Lawrence Berkeley National Laboratory, Berkeley, CA 94720, USA}

\author{Aaron Bostwick}\email[abostwick@lbl.gov]{}\affiliation{Advanced Light Source, E. O. Lawrence Berkeley National Laboratory, Berkeley, CA 94720, USA}

\begin{abstract}

We investigate the atomic-scale tunneling characteristics of bilayer graphene on silicon carbide using the scanning tunneling microscopy. The high-resolution tunneling spectroscopy reveals an unexpected negative differential resistance (NDR) at the Dirac energy, which spatially varies within the single unit cell of bilayer graphene. The origin of NDR is explained by two near-gap van Hove singularities emerging in the electronic spectrum of bilayer graphene under a transverse electric field, which are strongly localized on two sublattices in different layers. Furthermore, defects near the tunneling contact are found to strongly impact on NDR through the electron interference. Our result provides an atomic-level understanding of quantum tunneling in bilayer graphene, and constitutes a useful step towards graphene-based tunneling devices. 

\end{abstract}
\pacs{73.22.Pr, 73.20.At, 74.55.+v, 68.37.Ef} 

\maketitle

Understanding quantum tunneling at the atomic level is essential in the study of nanoscale materials and their applications in the tunneling devices \cite{PONO,BRIT}. One of the most interesting tunneling phenomena is negative differential resistance (NDR), characterized by the reversal of the standard current-voltage relationship, decreasing current with increasing voltage. NDR is the basic operating principle of Esaki and resonant-tunneling diodes, and has enabled various novel applications \cite{ESK,RTD}. Motivated by its fundamental importance and potential applications, there have been efforts to study NDR in graphene \cite{ZFW,LAKE,LEV,FIORI,WU}, a prototypical two-dimensional material with tunable electronic properties \cite{GEIM,OHTA,WANG,AVO}. However, tunneling-induced NDR has not yet been realized in graphene, while a recent study has proposed a new method for NDR in graphene that is not based on the quantum tunneling effect \cite{WU}.

The scanning tunneling microscope (STM) is a powerful tool, which can not only directly probe NDR, but also provide key information on the mechanism. STM has been widely employed to study various electronic properties of graphene at the atomic scale, such as scattering and interference \cite{RUT1,KERN}, but investigation into the sub-unit cell regime has been limited. Here, we report the first observation of NDR based on quantum tunneling at the vertical junction of STM over bilayer graphene. The applied electric field across bilayer graphene induces two van Hove singularities in the electronic spectrum, which are strongly localized on two sublattices in different layers. Such a localization of electronic singularities leads to a novel atomic-scale variation of NDR, which is directly visualized by our high-resolution tunneling spectroscopy within the single unit cell of bilayer graphene. This result provides the atomic-level understanding of quantum tunneling in bilayer graphene.

Bilayer graphene was prepared on 6H-SiC(0001) wafers (N-dopant concentration of 1 $\times$ 10$^{18}$ cm$^{-3}$) by thermal graphitization in a flow of argon as described in Ref. \cite{EMT}. The samples were transferred through the air to the ultrahigh vacuum chamber (5 $\times$ 10$^{-11}$ torr), and briefly annealed up to 800 $^{\circ}$C to refresh the surface. STM measurements were performed at 5.6 K and 78.2 K using a commercial cryogenic STM (Unisoku, Japan), equipped with a Nanonis controller (Specs, Germany). {\dIdV} spectra were obtained by the standard lock-in technique with a modulation voltage of 3--18 mV at 500 Hz and a consistent setpoint of --0.2 V and 300 pA. The {\dIdV} map was taken by a grid spectroscopy, for which the thermal drift was carefully compensated by the atom-tracking method. 

%--- Figure 1 ---
\begin{figure}
\center
\includegraphics[scale=1]{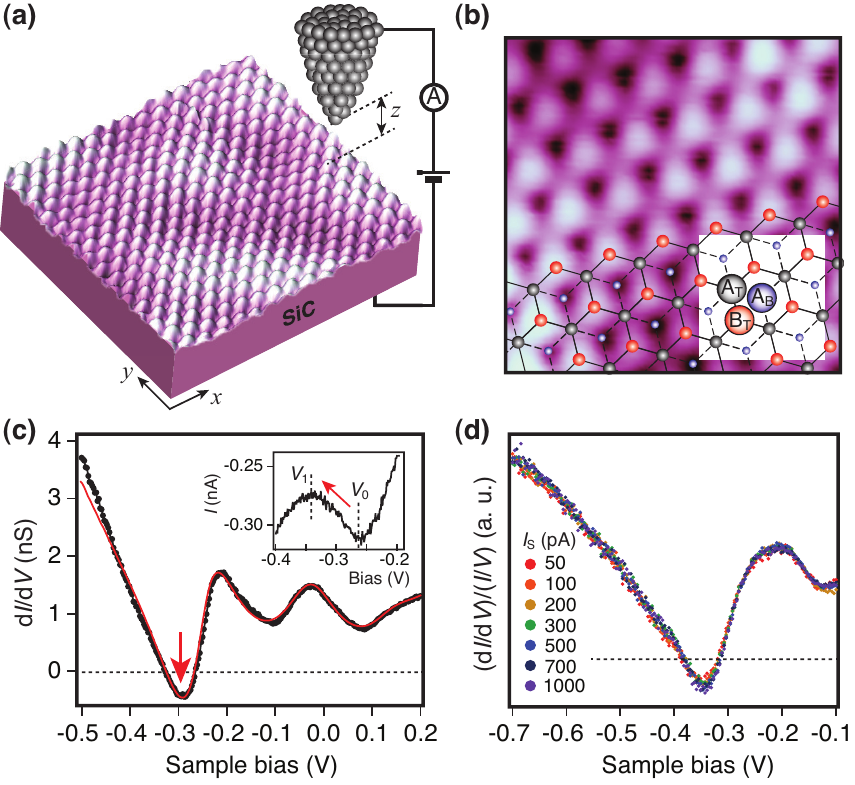}
\caption{(color online). (a) Experimental configuration for the local tunneling spectroscopy, consisting of an STM tip and bilayer graphene on silicon carbide. (b) STM topographic image (1.5 $\times$ 1.5 nm$^{2}$) taken on bilayer graphene with sample bias of --0.2 V and tunneling current of 300 pA. Inset shows the atomic structure overlaid with part of the STM image. Balls are carbon atoms, and solid (dashed) lines represent chemical bonds in the top (bottom) layer. (c) Point-{\dIdV} spectrum and a corresponding {\IV} spectrum (inset) taken on the {\BT} site. The red line overlaid is the result of model simulations. Red arrows highlight NDR. (d) {\dIdV} spectra normalized by {\IV} at each initial setting current marked with different colors. Dotted lines in (c) and (d) indicate conductance zero.}
\label{Fig1}
\end{figure}

The basic configuration of our experimental setup is shown in Fig. 1(a). It involves the tunnel junction of an STM tip and bilayer graphene separated by a vacuum gap of about 1 nm. Using this technique we can measure {\IV} curves over a specific atomic site in the surface. The bilayer graphene is \textit{n}-doped with {\EF} about 0.3 eV above the charge-neutrality point [the Dirac energy ({\ED})]. The interface of bilayer graphene with SiC(0001) produces a built-in electric field, rendering the charge densities in two stacked graphene layers inequivalent \cite{FALKO}. This out-of-plane symmetry breaking results in a bandgap with the magnitude of 0.14 eV at {\ED} \cite{OHTA}. This characteristic of bilayer graphene has enabled a tunable bandgap either by gating \cite{WANG} or chemical doping \cite{OHTA}, and is also important for NDR as explained below.

Figure 1(b) shows an STM topographic image of bilayer graphene, acquired with --0.2 V sample bias and constant tunneling current. The atomic structure of bilayer graphene is shown in the inset and overlaid with part of the image for comparison. Bilayer graphene consists of two Bernal stacked graphene layers, where the \textit{A} sublattice of the top layer ({\AT}, black) is located on the \textit{B} sublattice of the bottom layer ({\BB}). Due to the direct coupling between them, the other sublattice in the top layer ({\BT}, red) is dominantly imaged at low-bias voltage, resulting in the triangular pattern \cite{TERS,LAU}. The full honeycomb lattice could be visualized at the higher bias voltage, confirming the characteristic bias dependence of bilayer graphene \cite{RUT2,SUP}. In Fig. 1(b), one can also readily distinguish the {\AT} site from the hollow site in the top layer (corresponding to {\AB}, blue) by their relative brightness. This demonstrates the high spatial resolution achieved in our measurements.

Figure 1(c) shows a {\dIdV} curve measured on the {\BT} site. The differential conductance value goes below zero at --0.3 V, indicative of negative conductance (or resistance). This can also be directly seen in {\IV} measurements [inset of Fig. 1(c)] as decreasing current while increasing the voltage from --0.26 to --0.34 V. This observation clearly demonstrates NDR in bilayer graphene. The NDR behavior was reproducibly observed in the voltage range of --0.30 $\sim$ --0.35 V, close to {\ED}. In Fig. 1(d), we display a series of {\dIdV} spectra normalized by \textit{I}/\textit{V} taken with different initial tunneling  currents, which are inversely related to the tip-sample distance (tunneling barrier width). They show little change and consistent NDR behaviors, excluding tip-induced origins of NDR such as a local band bending \cite{JUNG}.

%--- Figure 2 ---
\begin{figure}
\includegraphics[scale=1]{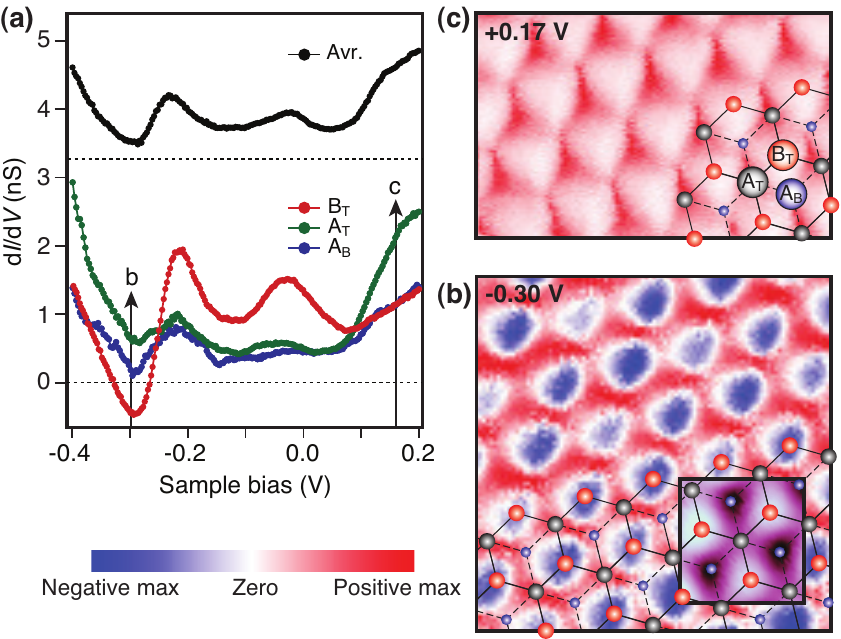}
\caption{(color online). (a) Point-{\dIdV} spectra taken on three different atomic sites of bilayer graphene, and its spatial average offset for clarity. Dotted lines indicate conductance zero. (b),(c) {\dIdV} maps taken with a 96 $\times$ 96 grid at the bias voltage indicated by arrows in (a). The inset in (b) shows the simultaneously acquired topographic image overlaid with the atomic structure of bilayer graphene.}
\label{Fig2}
\end{figure}

An important clue for the origin of NDR can be obtained from its spatial distribution at the atomic scale. Figure 2(a) compares {\dIdV} curves on three different atomic sites [refer to Fig. 1(b)]. NDR at --0.3 V occurs only on the {\BT} site, while above +0.1 V the dominant conductance is found on the {\AT} site. Recording {\dIdV} curves as a function of lateral tip positions, we show a differential conductance map at --0.3 V in Fig. 2(b). The simultaneously acquired topographic image in the inset helps determine the atomic position. Interestingly, negative conductance, represented by blue color, is centered on each {\BT} atom and is surrounded by a hexagonal rim of positive conductance. This suggests that NDR is closely linked to localized states at the {\BT} site. On the other hand, the same map taken at +0.17 V shows dominant positive conductance on the {\AT} site. This site-resolved detail has not been observed previously most likely due to limited spatial resolution in the tunneling spectroscopy. Indeed, the spatial average of {\dIdV} curves [the black one in Fig. 2(a)] reproduces the typical curve with no NDR reported previously in literature \cite{LAU}.

%--- Figure 3 ---
\begin{figure}
\includegraphics[scale=1]{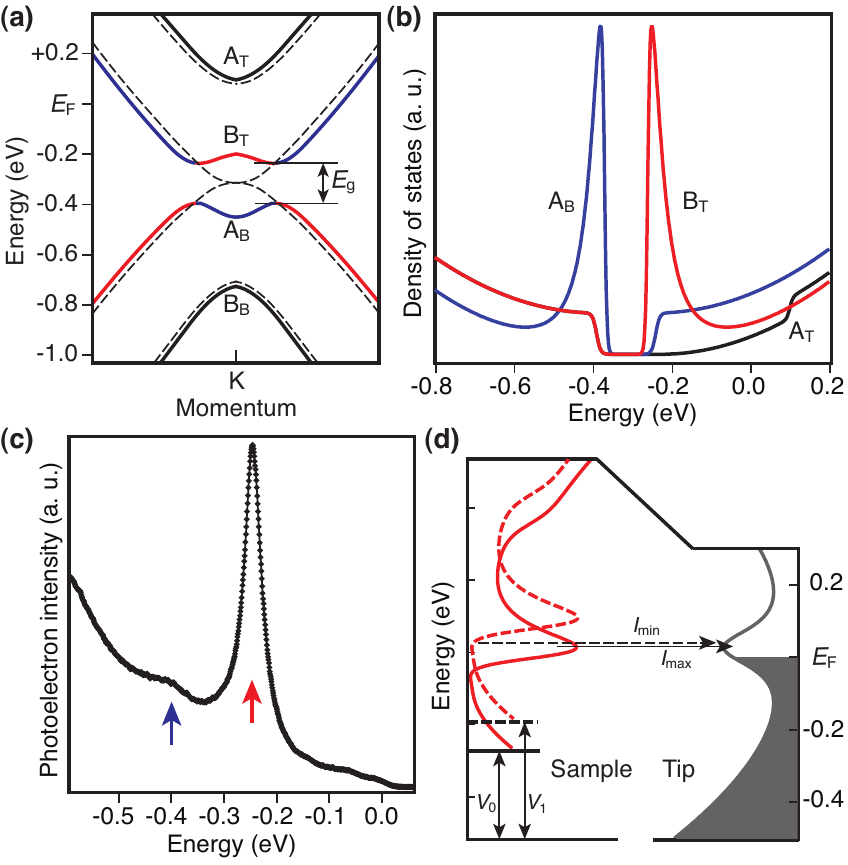}
\caption{(color online). (a) Band structure of bilayer graphene with (solid lines) and without (dashed lines) the bandgap. Red and blue colors in a low-energy band pair represent the atomic site, at which corresponding states are localized ({\BT}: red, {\AB}: blue). (b) Site-resolved density of states in biased bilayer graphene \cite{YU,TOWE}. (c) Energy-distribution curve at the K point taken by ARPES. Data were taken at the Advanced Light Source, using a Scienta R4000 electron analyzer (VG-Scienta, Sweden) and a 95-eV photon. Energy and angular resolutions were 30 meV and 0.1$^{\circ}$. Red and blue arrows indicate positions of two van Hove singularities. (d) Energy diagram for the tunnel junction of the tip and the sample. Red solid and dashed lines are, respectively, the sample density of states at \textit{V}$_{0}$ for local maximum current (\textit{I}$_{\mathrm{max}}$) and at \textit{V}$_{1}$ for local minimum current (\textit{I}$_{\mathrm{min}}$) after broadening by Gaussian. The gray line is the tip density of states, and the shaded region is the filled tip states below {\EF}.}
\label{Fig3}
\end{figure}

To understand the origin of NDR, let us recall the band structure of bilayer graphene near {\EF} as illustrated in Fig. 3(a). In pristine bilayer graphene, there are two pairs of energy bands (dotted lines) --- a low-energy pair from the {\BT} and {\AB} atoms and a high-energy pair from the {\AT} and {\BB} atoms. When a bandgap is induced by a transverse electric field, the low-energy bands develop a Mexican-hat-like dispersion at conduction and valence band edges near the bandgap \cite{MCFAL,MIN}. This rather flat dispersion leads to two distinct van Hove singularities in the density of states \cite{GUI} as in Fig. 3(b). Furthermore, these two near-gap states were theoretically predicted to be localized at specific atomic sites in different layers, as shown in Figs. 3(a) and 3(b), depending on the direction of the electric field \cite{YU,TOWE}. Such a strong localization of electronic singularities (the flat parts of energy bands) is also important in understanding many-body correlations at low energies and the ground state of bilayer graphene \cite{VELA}, but has not been directly shown by experiments.

In our case, states near --0.25 and --0.39 eV are localized at {\BT} in the top layer (red) and at {\AB} in the bottom layer (blue), respectively, while states near +0.1 eV are localized at {\AT} in the top layer (black). Since STM is highly sensitive to the local density of states in the surface, this explains dominant conductance at {\BT}  for --0.2 $\sim$ +0.1 V in Fig. 2(a) and that at {\AT} above +0.1 V in Figs. 2(a) and 2(c). This picture is fully supported by the energy-distribution curve at the K point [Fig. 3(c)], measured by angle-resolved photoemission spectroscopy (ARPES). It reveals highly asymmetric intensities for the two van Hove singularities \cite{OHTA}, because the photoelectrons excited from the bottom layer ({\AB}) are attenuated by scattering during their passage though the top layer. This also confirms that the layer localization is universal over the wide area covered by the photon beam ($\sim$50 $\mu$m in radius) in ARPES.

The presence of a layer-polarized van Hove singularity can qualitatively explain NDR, as in Fig. 3(d), which shows the relationship between the broadened sample density of states at {\BT} and the tip density of states. When the sample bias equals \textit{V}$_{0}$, where the van Hove singularity meets the {\EF} of the tip, the current reaches a local maximum, \textit{I}$_{\mathrm{max}}$. As it approaches \textit{V}$_{1}$, where the van Hove singularity no longer assists tunneling and no state exists in the bandgap, tunneling is strongly suppressed, and the current decreases down to \textit{I}$_{\mathrm{min}}$. As the voltage further increases out of the bandgap, the current starts increasing again, resulting in the N-shaped branch in the {\IV} curve [as in the inset of Fig. 1(c)]. That is, the observed NDR is the result of electronic spectrum reconstructions in asymmetric bilayer graphene.

For the quantitive analysis, we performed {\IV} simulations based on the standard formula, \textit{I}(\textit{V},\textit{z}) $\propto$ $\int$$^{\textit{E}_{\mathrm{F}}}_{\textit{E}_{\mathrm{F}}-V}$DOS$_{\mathrm{sample}}$(\textit{E}+\textit{V})$\cdot$DOS$_{\mathrm{tip}}$(\textit{E})$\cdot$\textit{T}(\textit{E},\textit{V},\textit{z})\textit{dE}, where DOS is the density of states and \textit{T} is tunneling matrix element \cite{CROM,LYO,LIMOT}. \textit{T} is estimated by the WKB approximation as \textit{T}(\textit{E},\textit{V},\textit{z}) = exp(--2$\sqrt{2(\phi-V/2-E)}$\textit{z}, where $\phi$ is the work function. We used the Gaussian-broadened local DOS in the theoretical calculation \cite{YU,TOWE} at {\BT} with a small contribution of that at {\AT} (to take into account the finite areal averaging). The tip DOS is assumed to have a typical localized state of tip-apex atoms near {\EF} \cite{LYO}. The sample and tip DOS's optimized to reproduce the experimental data are present, respectively, by red and gray curves in Fig. 3(d). The simulated curve [the red line overlaid in Fig. 1(c)] successfully reproduces the NDR behavior, confirming the proposed mechanism.

%--- Figure 4 ---
\begin{figure}
\includegraphics[scale=1]{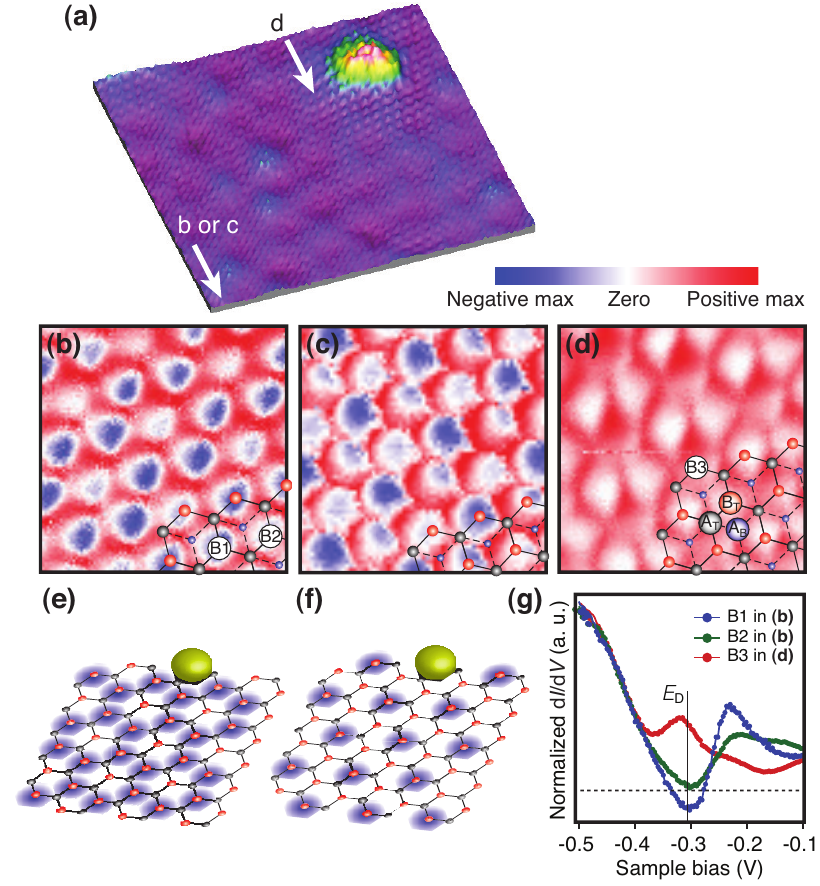}
\caption{(color online). (a) STM image (10 $\times$ 10 nm$^{2}$) with --0.05 V and 300 pA, where a local defect (the yellow blob) is present at the corner. {\dIdV} maps taken (b),(c) at two different regions about 16 nm away from defects and (d) at a region right next to a defect. The atomic structure of bilayer graphene is overlaid. (e),(f) Schematic illustration for two possible NDR patterns due to constructive or destructive interference of defect scatterings. (g) Normalized {\dIdV} spectra taken on three different {\BT} atoms marked in (b) and (d). Solid and dotted lines indicate {\ED} and conductance zero, respectively.}
\label{Fig4} 
\end{figure}

To further support the proposed mechanism, we have investigated the interaction of NDR with local defects. Recent theories predict that a single point scatterer in bilayer graphene develops in-gap states, suppressing the sharp van Hove singularities \cite{NETO, PUCC}. Thus, one might expect a local suppression of NDR near defects. Figure 4(a) shows a large scale STM image of an impurity (yellow blob). The {\dIdV} maps taken at two different locations about 16 nm away from defects are present in Figs. 4(b) and 4(c). NDR is extinguished in a subset of {\BT} atoms, which form a periodic {\R3byR3} pattern rotated by 30$^{\circ}$ with respect to the lattice constant, and these data show little dependence on the origin and site of defects. The ({\R3byR3})R30$^{\circ}$ modulation is the well-known signature of intervalley electron scattering by a defect in bilayer graphene \cite{RUT1,KERN}. Thus, we attribute two inverse patterns in Figs. 4(b) and 4(c) to constructive or destructive interference of defect scattering, as illustrated in Figs. 4(e) and 4(f). Furthermore, the same map taken right next to a defect [Fig. 4(d)] shows that NDR completely disappears and instead strong positive conductance emerges on a subset of {\BT} atoms. As a result, positive, zero, and negative conductance are all possible [solid line in Fig. 4(g)], and defects can be used to switch NDR on and off nearby as well as at the site more than 10 nm away from a defect. Such a strong impact of defects in electronic localization should be considered for the complete understanding of scattering and interference in bilayer graphene \cite{RUT1,KERN}. The disappearance of NDR with the progressive suppression of the van Hove singularity near the bandgap [the peak at --0.23 V in Fig. 4(g)] makes a clear link between them. This will further allow the control of NDR by tuning the bandgap with gating or chemical doping \cite{OHTA,WANG}.

Our results, combined with a few existing techniques, suggest a route towards graphene-based NDR devices. The key to NDR is, as demonstrated here, to exploit the near-gap van Hove singularity in biased bilayer graphene. In our experimental configuration, the tip can be replaced by another biased bilayer grahene, such that a vertical tunnel junction of two graphene bilayers is formed. A similar graphene heterostrucrture has recently been realized using ultrathin boron nitride films as a tunneling barrier \cite{PONO,BRIT}. In such junctions, tunneling would depend on sublattice-registry matching. That is, if the sublattices in two graphene layers facing each other, where the van Hove singularity is localized, were to line up, maximal direct tunneling between van Hove singularities could be achieved. The tunable electronic structure of bilayer graphene \cite{OHTA,EVA,SON} would allow further optimization for NDR and a large peak-to-valley ratio of tunneling current.

%--- Acknowledgements ---
This work and ALS were supported by the Director, Office of Science, Office of Basic Energy Sciences, of the U.S. Department of Energy under Contract No. DE-AC02-05CH11231. K.S.K. acknowledges support by NRF Grant (NRF-2011-357-C00022). Work in Pohang was supported by NRF through Center for Low Dimensional Electronic Symmetry (Grant No. 2012R1A3A2026380) and SRC Center for Topological Matter (Grant No. 2011-0030789). Work in Erlangen was supported by the DFG through grant number SE 1087/7-1, by the ESF Eurocores program EUROGRAPHENE, and by the DFG priority program 1459 ÒGrapheneÓ. A.L.W. acknowledges support by the Max Planck Society. We thank F. Speck, M. Ostler, and F. Fromm for assistance during sample preparation.

\newpage
%--- Figure S1 ---
\begin{figure*}
\includegraphics[scale=1]{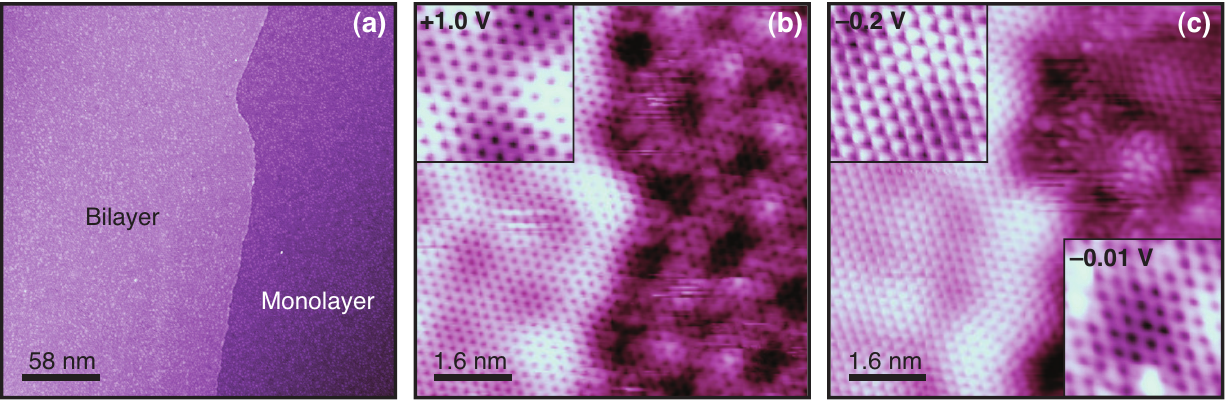}
\caption{(a) Large scale STM image taken with +2 V and 50 pA, where two wide domains of monolayer and bilayer graphene are clearly distinguished. (b),(c) STM images at the boundary of monolayer (right) and bilayer domains (left), taken with two different bias voltages marked at the upper left. Insets at the upper left show magnified images of the left domains. The full honeycomb pattern at high-bias voltage (b) transforms into the triangular pattern at low-bias voltage (c), which is the characteristic bias dependence in bilayer graphene reported previously [20]. This confirms the bilayer origin of domains under study. The lower right inset in (c) shows a magnified low-bias (marked) image of the right domains, showing a clear honeycomb lattice of monolayer graphene. }
\end{figure*}

\end{document}